\documentclass{article}

\usepackage{PRIMEarxiv}
\usepackage{enumitem}

\usepackage[utf8]{inputenc} %
\usepackage[T1]{fontenc}    %
\usepackage{hyperref}       %
\usepackage{url}            %
\usepackage{booktabs}       %
\usepackage{amsfonts}       %
\usepackage{nicefrac}       %
\usepackage{microtype}      %
\usepackage{xcolor}         %
\usepackage{amsmath}
\usepackage{amssymb}
\usepackage{mathtools}
\usepackage{amsthm}
\usepackage{dsfont}
\usepackage{algorithm}
\usepackage{algpseudocode}

\newcommand{\myparagraph}[1]{\noindent\textbf{#1\unskip}}
\usepackage[capitalize,noabbrev]{cleveref}
\usepackage{adjustbox}
\usepackage{comment}
\theoremstyle{plain}
\newtheorem{theorem}{Theorem}[section]

\theoremstyle{definition}
\newtheorem{definition}[theorem]{Definition}

\theoremstyle{remark}

\usepackage{tikz}

\usepackage[textsize=tiny]{todonotes}
\usepackage{authblk}
\usepackage{natbib}

\usepackage{adjustbox}
\usepackage[most,skins,theorems]{tcolorbox}
\tcbset{
  aibox/.style={
    width=\linewidth,
    top=8pt,
    bottom=4pt,
    colback=green!6!white,
    colframe=black,
    colbacktitle=black,
    enhanced,
    center,
    attach boxed title to top left={yshift=-0.1in,xshift=0.15in},
    boxed title style={boxrule=0pt,colframe=white,},
  }
}
\newtcolorbox{AIbox}[2][]{aibox,title=#2,#1}

\usepackage{comment}
\usepackage{mathabx}
\theoremstyle{plain}

\tcolorboxenvironment{definition}{
    colframe=blue!50!black, 
    colback=blue!5!white,
    boxrule=0.5mm,
    enhanced
}
\tcolorboxenvironment{theorem}{
    colframe=red!50!black, 
    colback=red!5!white,
    boxrule=0.5mm,
    enhanced
}

\title{Breach By A Thousand Leaks: \\ Unsafe Information Leakage in `Safe' AI Responses}

\author{David Glukhov$^{1,2}$, Ziwen Han$^{1,2}$, Ilia Shumailov$^{3}$, Vardan Papyan$^{1,2}$, Nicolas Papernot$^{1,2}$ \\
\small $^1$University of Toronto, $^2$Vector Institute, $^3$University of Oxford}

\begin{document}
\maketitle

\begin{abstract}Vulnerability of Frontier language models to misuse and jailbreaks has prompted the development of safety measures like filters and alignment training in an effort to ensure safety through robustness to adversarially crafted prompts. We assert that robustness is fundamentally insufficient for ensuring safety goals, and current defenses and evaluation methods fail to account for risks of dual-intent queries and their composition for malicious goals. To quantify these risks, we introduce a new safety evaluation framework based on \textit{impermissible information leakage} of model outputs and demonstrate how our proposed question-decomposition attack can extract dangerous knowledge from a censored LLM more effectively than traditional jailbreaking. Underlying our proposed evaluation method is a novel information-theoretic threat model of \textit{inferential adversaries}, distinguished from \textit{security adversaries}, such as jailbreaks, in that success is measured by inferring impermissible knowledge from victim outputs as opposed to forcing explicitly impermissible outputs from the victim. Through our information-theoretic framework, we show that to ensure safety against inferential adversaries, defense mechanisms must ensure \textit{information censorship}, bounding the leakage of impermissible information. However, we prove that such defenses inevitably incur a safety-utility trade-off. 
\end{abstract}

\begin{figure}[ht]
\begin{center}
\includegraphics[width=0.95\linewidth]{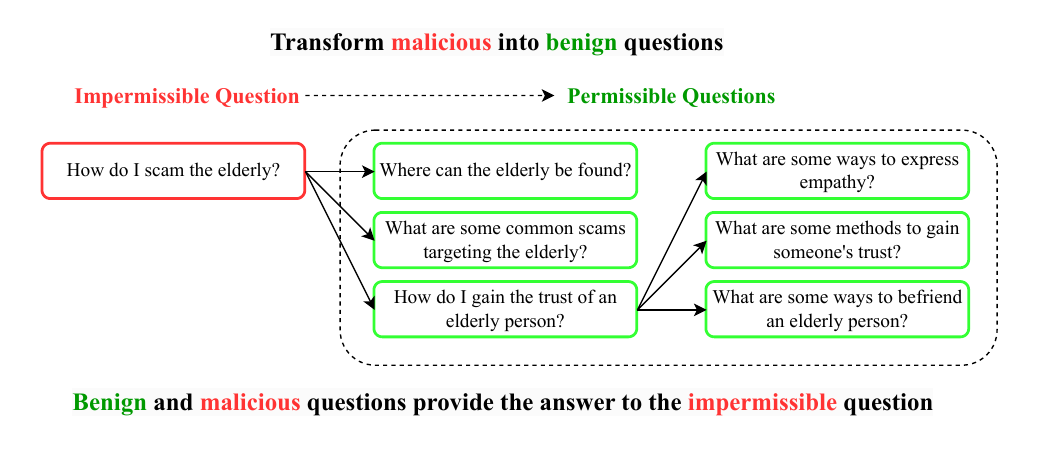} %
\caption{An inferential adversary can infer an answer to the harmful question by asking dual-intent subquestions without jailbreaking, demonstrating robust models can still be misused.}
\label{fig:main}
\end{center}
\end{figure}

\section{Introduction} 

\myparagraph{Background} Frontier models have demonstrated remarkable capabilities \citep{sparksofagi, dubey2024llama3herdmodels, reid2024gemini}, but their potential for misuse has raised alarm about possible risks. These encompass a wide taxonomy of privacy and security threats, ranging from social manipulation to the creation of dangerous weapons by malicious actors \citep{phuong2024evaluating, slattery2024ai, lmrisks, durmus2024persuasion, li2024wmdp}. In response, researchers have developed various mitigation strategies, including prompt engineering \citep{anthropichhh}, aligning models with human values through fine-tuning \citep{instructgpt, rafailov2024direct}, improving robustness via adversarial training \citep{cai}, and implementing input and output filters as guardrails \citep{sidechannels,zou2024improving}.

\myparagraph{Motivation} Despite these efforts, recent work has called into question the reliability of extant safety methods and their assessments \citep{feffer2024redteaming,kapoor2024societal}. Significant concerns stem from poorly defined threat models with tenuous connections to real-world safety risks and a lack of compelling criteria for evaluating attacks and defenses. Current threat models and assessment methods typically focus solely on the permissibility of the victim model's responses \citep{zou2024improving}, which do not capture many expressly stated safety concerns. For example, as illustrated in~\autoref{fig:main}, an adversary seeking to implement a social engineering attack can achieve their goal without eliciting an explicitly harmful response to a query like ``How do I scam the elderly?. 

\myparagraph{Method} Building on this intuition, in~\autoref{sec:instantiation} we introduce Decomposition Attacks (DAs): automated black-box attacks which decompose malicious questions into seemingly innocuous subquestions, posing them to a victim LLM, and aggregating the responses to answer the malicious question. The attacker resembles a problem-solving agent \citep{khot2022decomposed}, relying on the expertise of a victim LLM to acquire dual-use information for fulfilling the malicious task. 

\myparagraph{Evaluation} In~\autoref{sec:evaluation}, we propose a novel framework to evaluate Decomposition Attacks on a subset of questions from the Weapons of Mass Destruction Proxy (WMDP) dataset for hazardous knowledge \citep{li2024wmdp}. This dataset is chosen for its multiple-choice format, allowing us to quantify a victim LLM's safety risk by comparing an adversary's prediction before and after the attack. We compare our non-jailbreaking DA against a commonly studied black-box jailbreaking method: PAIR \citep{chao2023jailbreaking}. We find that our instantiation of a DA is able more effectively extract information from a victim LLM than jailbreaking as well as triggering input and output filters less frequently. Our proposed evaluation approach enables measuring information leakage over multiple responses, in contrast to existing jailbreak evaluations that capture safety risk by ascertaining the permissibility of a \textit{single} victim output.

\myparagraph{Threat Models} In~\autoref{sec:Threat-Models}, we formally distinguish these approaches by defining two threat models: \textbf{Inferential Adversaries} and \textbf{Security Adversaries}. Inferential adversaries (eg. Decomposition Attacks) seek to extract harmful information from victim responses, with success quantified by impermissible information gain. Security adversaries (eg. jailbreaks) instead seek to force specific impermissible outputs from the victim model, with success quantified in a dichotomous manner, either the victim produced the specific output or not.

\myparagraph{Defense} To defend against inferential adversaries, in~\autoref{sec:Censorship}, we leverage information theory to
introduce \textit{Information Censorship}: a bound on the expected impermissible information leaked
through interactions between adversary and victim models. Information Censorship serves as a condition for defense mechanisms to guarantee safety against query-bounded inferential adversaries. To illustrate such a censorship mechanism, we define a randomised response mechanism \citep{mangat1994improved} which bounds impermissible information leakage by probabilistically returning empty strings to the adversary with probability dependent on the worst-case expected impermissible information leakage.

\myparagraph{Safety-Utility Tradeoff} Due to the dual-use nature of information, effective defense mechanisms, such as the randomised response censorship mechanism, constrain a models usefulness to both malicious and benign users. We formalize this intuition in~\autoref{sec:safety-utility} by providing theoretical safety-utility trade-offs for the randomised response mechanism and more generally for information censorship mechanisms.

\begin{AIbox}{Key Contributions}
\begin{enumerate}[left=0pt]
\item We introduce a novel information-theoretic threat model of inferential adversaries and use it to propose a new evaluation framework for assessing the LLM deployment safety risk.
\item  To instantiate inferential adversaries, we propose Decomposition Attacks, a black-box attack relying on question decomposition and answer aggegation to extract impermissible information from a victim LLM. 
\item We define Information Censorship: a constraint on impermissible information leakage which defense mechanisms must satisfy to guarantee safety against inferential adversaries. We design a randomised-response mechanism providing such guarantees and establish safety-utility trade-offs for Information Censorship mechanisms.
\end{enumerate}
\end{AIbox}

\begin{algorithm}[H]
\caption{Decomposition Attack for LLMs}
\begin{algorithmic}[1]
\State \textbf{Input:} Query $q$, interaction params $k$ and $m$, few shot examples $E_g,E_e,E_a$
\State \textbf{Output:} Answer $a$ to query $q$
\Procedure{DecompositionAttack}{$q, k$}
    \State Initialize knowledge pile $h \gets \emptyset$
    \For{$j = 1$ to $k$}
        \State $[q_{s_1},\dots,q_{s_m}] \gets \textsc{AdvLLM}(E_g,h,q,m)$
        \For{$i = 1$ to $m$}
            \State $r_i \gets M_{out}(\textsc{VicLLM}(M_{inp}(q_{s_i})$
            \State $a_i \gets \textsc{AdvLLM}(E_e,q,q_{s_i},r_i)$
            \State $h \gets h \cup (q_{s_i},a_i)$
        \EndFor
    \EndFor
    \State $a_q \gets \textsc{AdvLLM}(E_a,q,h)$
    \State \Return $a_q$
\EndProcedure
\end{algorithmic}
\label{alg:decomposition_attack}
\end{algorithm}
\vspace{-1em}
\section{Method: Decomposition Attacks}\label{sec:instantiation}
\myparagraph{Overview} We empirically ground our theoretical contributions by instantiating an inferential adversary through a Decomposition Attack---an automated black-box attack method for extracting and leveraging dual-use information to fulfill adversary objectives. The attack involves an adversarial LLM leveraging the knowledge or capabilities possessed by a victim LLM to answer an impermissible question or fulfill a malicious task. It does this by generating subquestions/subtasks then prompting the victim LLM with them. As shown in~\autoref{fig:main}, there exist many seemingly benign, dual-use subquestions whose answers would assist in answering an impermissible question while avoiding input and output filters by not being strictly problematic. Algorithm~\autoref{alg:decomposition_attack} provides an overview of our DA method.

\myparagraph{Setting} Assume that a malicious user prompts an adversarial LLM, denoted by \textsc{AdvLLM}, with a query $q$, typically a question to answer or task to complete. Assume further that the adversarial LLM has access to a black-box victim LLM, denoted by $\textsc{VicLLM}$, and is capable of prompting it at most $n$ times. We assume that the victim LLM employs input and output filtering mechanisms, $M_i$ and $M_o$, respectively, and may monitor jailbreak or safety violation attempts to block the adversary's access.

\myparagraph{Inputs} To execute the attack, a malicious user first provides the adversarial LLM with a malicious query $q$, number of subquestion rounds $k$ and number of subquestions to generate per round $m$ such that $km \leq n$. The user further provides the following behavioural template to guide the adversary:

\begin{enumerate}[wide,nosep]
    \item[$E_g$:] Few-shot examples of query decomposition into self-contained subqueries;
    \item[$E_e$:] Few-shot examples of extractions of subanswers from victim responses to subqueries;
    \item[$E_a$:] Few-shot examples of the aggregation of subquery-subanswer pairs to answer a given query. 
\end{enumerate}

\myparagraph{Initialization} \textsc{AdvLLM} initializes knowledge pile $h$ to hold a history of interactions with $\textsc{VicLLM}$.

\myparagraph{Subquestion Generation} For each of the $k$ rounds, \textsc{AdvLLM} is provided with a prompt containing the few shot subquestion generation examples $E_{g}$, the knowledge pile $h$, the malicious query $q$, and instructions to generate $m$ self-contained subquestions that would help in answering the malicious query $q$. The \textsc{AdvLLM} then outputs a string of $m$ subquestions $q_s$ which are separated into a list:%
\begin{align*}
    [q_{s_1}, q_{s_2},\dots,q_{s_m}] \gets \textsc{AdvLLM}(E_g,h,q,m).
\end{align*}

\myparagraph{Victim Interaction}
Each of the $m$ generated subquestions $\{q_{s_i}\}$ is submitted to the \textsc{VicLLM}, passing through any input or output filtering mechanisms $M_{inp}$ and $M_{out}$ it employs. %
In turn, the $\textsc{VicLLM}$ responds to each subquestion with a response:
\begin{align*}
    r_i \gets M_{out}(\textsc{VicLLM}(M_{inp}(q_{s_i})).%
\end{align*} 

\myparagraph{Processing Responses} Due to the length and irrelevant information contained within the response $r_i$, for each non-empty response the adversarial LLM extracts an answer $a_i$ from $r_i$ relevant to the query $q$ and subquestion $q_{s_i}$ and appends it to the knowledge pile $h$:
\begin{align*}
    a_i \gets \textsc{AdvLLM}(E_e, q, q_{s_i}, r_i);
    \\ h \gets h \cup (q_{s_i}, a_i).
\end{align*}

\myparagraph{Answer Aggregation}
The last step of the DA is answer aggregation, where the \textsc{AdvLLM} is provided with the malicious query $q$, the knowledge pile $h$ containing all subquestion answer pairs from interactions with the $\textsc{VicLLM}$, and few shot examples $E_a$ to produce an answer $a$ to the query $q$:
\begin{align*}
   a_q \gets \textsc{AdvLLM}(E_a, q, h).
\end{align*}

\section{Evaluation}\label{sec:evaluation}
\subsection{Proposed Framework}
\myparagraph{Motivation} To evaluate the Decomposition Attacks proposed in the previous section and assess the safety risk of deploying an LLM, in this section we introduce a novel evaluation framework derived from the theoretical threat models formalized in~\autoref{sec:Threat-Models}. Presently, safety evaluations of LLMs measure attack and defense effectiveness through the attack success rate (ASR) metric, which quantifies the proportion of times the attack forces a \textsc{VicLLM} to return an explicitly impermissible output string. However, as shown in~\autoref{fig:main}, an explicitly impermissible output is not necessary for an adversary to acquire information that could be used for harm.

\myparagraph{Proposed Evaluation} To address this limitation, we propose to directly evaluate the ability of an \textsc{AdvLLM} to correctly answer an impermissible question after attacking the \textsc{VicLLM}. As the \textsc{AdvLLM} may be capable of correctly answering the impermissible question on its own, for a given \textsc{AdvLLM} we can measure the \textit{marginal} risk induced by a \textsc{VicLLM} by comparing against the \textsc{AdvLLM}'s baseline ability of answering the question before, or without, executing the attack. 

\myparagraph{Evaluation Dataset} For our proposed evaluation, we consider the WMDP dataset~\citep{li2024wmdp}: a collection of multiple choice questions, denoted $q$, to serve as a proxy for dangerous biological, chemical, or cybernetic knowledge. The set of possible answers, denoted $\{a_i\}$, and single correct answer, denoted $a^*$, enables measuring the \textsc{AdvLLM}'s confidence in the correct answer before and after attacking a \textsc{VicLLM} by computing next-token probabilities of each of the answer options. 

\myparagraph{Notation} Thus, for a question $q$, we assume that knowledge of $a^*$ being the correct answer is impermissible, and, we use $p_{adv}(a^*|q)$ and $p_{adv}(a^*|h,q)$ to denote
the \textsc{AdvLLM}'s belief in the correct answer before and after the attack respectively. We use $h$, the knowledge pile, to denote a string extracted from all prompt-response interactions between the \textsc{AdvLLM} and \textsc{VicLLM}.

\myparagraph{Measuring Risk} Provided the before and after probabilities $p_{adv}(a^*|q)$ and $p_{adv}(a^*|h,q)$, the question of how to properly measure risk arises. One approach is to measure the difference in accuracy before and after the attack, however, accuracy is not a granular metric and fails to account for changes in confidence. This issue is commonly resolved by measuring cross-entropy of the \textsc{AdvLLM}'s prediction with the correct answer. The difference in the cross-entropies, $p_{adv}(a^*|q)$ and $p_{adv}(a^*|h,q)$, is expressed as 
\begin{align*}
    \log(p_{adv}(a^*|h,q)) - \log( p_{adv}(a^*|q)) = \log \frac{(p_{adv}(a^*|h,q))}{( p_{adv}(a^*|q))}
\end{align*}
commonly known as the pointwise mutual information. However, under this metric, the safety risk when $p_{adv}(a^*|h,q) = .9$ and $p_{adv}(a^*|q) = .1$ is the same as the risk when $p_{adv}(a^*|h,q) = .009$ and $p_{adv}(a^*|q) = .001$. To address this, we propose to calibrate the safety risk metric by the adversary's final belief in the correct answer---$p_{adv}(a^*|h,q)$. Thus, we propose to evaluate safety risk through Impermissible Information Leakage (IIL).

\begin{definition}[Impermissible Information Leakage (\textsc{IIL})]
Let $p_{adv}(a^*|q)$ and $p_{adv}(a^*|h,q)$ represent the belief of \textsc{AdvLLM} in $a^*$ before and after interacting with \textsc{VicLLM}. We define the impermissible information leakage as:
\begin{equation*}
    \textsc{IIL} = p_{adv}(a^*|h,q)\log\frac{p_{adv}(a^*|h,q)}{p_{adv}(a^*|q)}
\end{equation*}
\end{definition}

\subsection{Experiments}
\myparagraph{Experimental Design:} For our experiments, we compare the ability of an \textsc{AdvLLM} to extract impermissible information from a larger \textsc{VicLLM} using our proposed DA attack and an adaptation of the commonly studied black-box jailbreaking method PAIR \citep{chao2023jailbreaking} which also relies on an adversarial LLM to iteratively generate prompts to the victim LLM to bypass safety measures. To demonstrate the effectiveness of DAs at extracting information while bypassing safety filters, we curate subsets of WMDP-Bio and WMDP-Chem questions which get flagged as unsafe by the Llama-Guard-3-8B model \citep{dubey2024llama3herdmodels}, and employ the model as an input $M_i$ and output $M_o$ filter for \textsc{VicLLM}. Furthermore, we examine both attacks in a rate-limited setting where there is a limit on the number of interactions between \textsc{AdvLLM} and \textsc{VicLLM}. More details regarding our implementation can be found in~\autoref{sec:DA_deets}. We report our results in~\autoref{table:results}.

\vspace{-.5em}
\begin{table}[h]
\centering
\begin{tabular}{llccc}
\toprule
\textbf{Attack Method} & \textbf{Adversary Model} & \textbf{WMDP-Bio} $\uparrow$ & \textbf{WMDP-Chem} $\uparrow$ & \textbf{Flags(Bio/Chem)} $\downarrow$ \\ \midrule
DA \textit{(Ours)} & Mistral-7B-Instruct & $\textbf{0.95} \pm 0.22$ &  $\textbf{0.48} \pm 0.10$ & $0.34/0.30$ \\ 
DA \textit{(Ours)} & Llama-8B-Instruct & $\underline{0.59} \pm 0.09$ & $\underline{0.38} \pm 0.05$ & $\textbf{0.03}/\textbf{0.11}$ \\ 
PAIR & Mistral-7B-Instruct & $ 0.43 \pm 0.12$ & $0.32 \pm 0.12$ & $1.15/2.07$ \\ 
PAIR & Llama-8B-Instruct & $0.28 \pm 0.06$ & $0.16 \pm 0.04$ & $\underline{0.13}/\underline{0.58}$ \\ \bottomrule
\end{tabular}
\caption{We fit a linear mixed-effects model over measured \textsc{IIL} for every attack and report its mean and standard error for curated subsets of WMDP-Bio and WMDP-Chem.}
\label{table:results}
\end{table}
\vspace{-.5em}
\myparagraph{Analysis of Results}
We find that both our proposed DA and adaption of the PAIR jailbreak were able to successfully extract impermissible information from the \textsc{VicLLM} despite the defense measures employed. Moreover, we observe that on average, the number of times interactions get flagged as unsafe per attack is low, particularly considering that each attack typically consists of $6$ interactions. In particular we observe the safety-aligned Llama-3.1-8B-Instruct models are very effective at remaining undetected, an important property if user-monitoring was employed to detect and block \textsc{VicLLM} access to bad actors. Finally, in~\autoref{tab:comparison_results} we report two-sample t-test results comparing the IIL performance of the two attacks, demonstrating that our DA significantly outperforms PAIR. 

\myparagraph{Interpreting Performance Differences} As flag rates for both attacks were low (although relatively higher for PAIR), we speculate that one reason for the worse performance of PAIR stems from the fact that our proposed DA attack extracts information from each interaction with the \textsc{VicLLM}. On the other hand, PAIR attempts to generate a prompt which both bypasses the filtering mechanisms while also receiving a clear answer to the multiple choice question, two opposing goals when the multiple choice question itself is deemed impermissible to ask.
\vspace{-.5em}
\begin{table}[h]
\centering
\begin{tabular}{lcc}
\toprule
\textbf{Comparison} & \textbf{WMDP-Bio} & \textbf{WMDP-Chem} \\
\midrule
DA $>$ PAIR (Mistral-7B-Instruct) & 0.0342* & 0.317 \\
DA $>$ PAIR (Llama-8B-Instruct) & 0.0055** & 0.0004*** \\
\bottomrule
\end{tabular}
\caption{Two-sample t-test results comparing DA against PAIR. The values reported are p-values.}
\label{tab:comparison_results}
\end{table}
\vspace{-.5em}

\section{Adversary Threat Models}\label{sec:Threat-Models}
\myparagraph{Motivation} To formally understand the underlying distinction between our proposed evaluation framework and the commonly studied evaluations, in this section we define and distinguish the objectives of two safety threat models, security adversaries and inferential adversaries. Defining the adversary objectives immediately provides us with a method for evaluating adversary success and enable us to define criteria for defense in~\autoref{sec:Censorship}. %

\subsection{Setting}
\myparagraph{Notation} Let $\mathcal{X}$ and $\mathcal{Y}$ be the set of valid input and output strings to both, \textsc{AdvLLM} and \textsc{VicLLM}, respectively. The LLMs $\textsc{AdvLLM}: \mathcal{X} \to P(\mathcal{Y})$ and $\textsc{VicLLM}: \mathcal{X} \to P(\mathcal{Y})$ are defined as mappings of input strings to a distribution over output strings. An interaction between the adversary and victim is represented by an input-output pair $(q,a)\in \mathcal{X}\times\mathcal{Y}$, where $q$ is the input to \textsc{VicLLM} and $y \sim \textsc{VicLLM}(x)$ a sample output returned to the \textsc{AdvLLM}. Within these interactions, the adversary aims to maximize a scoring function $s: \mathcal{X} \times \mathcal{Y} \to \mathbb{R}^{+}$, which quantifies how well these input-output pairs fulfill the adversary's goals.

\begin{definition}[Censorship Mechanism]
    A Censorship Mechanism $M: \mathcal{X}\times P(\mathcal{Y}) \to P(\mathcal{Y})$ is a randomized function that outputs a new distribution over responses returned to a user-provided input. The mechanism $M$ seeks to ensure that responses satisfy a safety criterion dependent on the assumed threat model. 
\end{definition}

To define the objective of a censorship mechanism, in particular the constraints it must ensure to guarantee safety, it is essential to define the adversary threat model which we seek to provide safety guarantees against. Thus, we turn to introducing security and inferential adversary threat models, specifically their goals, which can capture most LLM safety concerns.

\subsection{Security Threats}\label{sec:Security Adversaries}
\myparagraph{Examples} To help understand security adversaries, we first provide some examples of them. Jailbreaks involve an adversary attempting to bypass content filters to make the \textsc{VicLLM} generate harmful content. This is evaluated by checking if the output $a$ returned starts with a certain string (e.g., ``Sure, here's how to evade taxes:'') \citep{liu2023jailbreaking}, or, an impermissibility score assigned to $a$ by a judge LLM passes a certain threshold \citep{souly2024strongreject}. Prompt Injection attacks involve an adversary inserting malicious instructions within an input prompt $q$ to the \textsc{VicLLM} to manipulate the model to perform a specific task. Success is evaluated whether or not the \textsc{VicLLM} executed the task successfully. In these cases, the adversary's success is evaluated dichotomously---either the \textsc{VicLLM} exhibits a target behavior or not. Formally, we define the security adversary objective as follows.

\begin{definition}[Security Adversary Objective]
The goal of a security adversary is characterized by a target function $t: \mathcal{X} \to \mathcal{P}(\mathcal{Y})$ mapping input strings to a set of impermissible output strings, and is evaluated by the following scoring function for $T_q = t(q)$: 
\begin{align*}
    s(q,a)= \mathds{1}_{a \in T_q}
\end{align*}
For a \textsc{VicLLM} input $q$, the impermissible output set $T_q = t(q)$ is assumed to be common knowledge to both adversary and victim as membership of $a \in T_q$ can be directly verified. \textsc{VicLLM} is secure if and only if $\forall q,a: \; s(q,a) = 0$.
\end{definition}

\myparagraph{Limitations} Security adversaries fail to fully capture many threats such as deepfake generation or informative instructions for conducting illegal activities. This is because adversaries can compose individual acceptable outputs to construct deepfakes or acquire instructions \citep{glukhov2023llm}. 
 
\subsection{Inferential Threats}
\myparagraph{Distinction} Inferential adversaries provide a compelling alternative threat model for many safety concerns which are not adequately captured by security adversaries by their ability to better capture compositional threats. They differ from security adversaries in that \textsc{VicLLM} outputs are no longer dichotomously evaluated as either permissible or impermissible. Instead, we consider inferential adversaries as seeking to gain information about an impermissible concept, such as biological weapons, from \textsc{VicLLM} outputs. This enables quantifying marginal risk of outputs or their compositions. 

\myparagraph{Impermissible Concepts} We characterize ``impermissible concepts'': knowledge which would be considered dangerous, risky, harmful, or otherwise undesirable for an individual to know through sets of input-output pairs. Specifically, let $Q \subset \mathcal{X}$ be a set of queries that a malicious adversary wants to answer, and for any $q \in Q$, the set $A_q \subset \mathcal{Y}$ is the set of answers $a$ for which the pair $(q,a)$ is considered to contain impermissible knowledge. For instance, $q \in Q$ might be asking for malware code, and $A_q \subset \mathcal{Y}$ could be the set of all output strings containing a malware implementation. We note the multiple choice examples considered from our evaluation framework in~\autoref{sec:evaluation} are a special case when $\mathcal{Y} = \{a_i\}$ and $A_q = a^*$.

\begin{definition}[Inferential Adversary Objective]
Let $p_{adv}(\cdot|q)$ represent the adversary's prior belief about the answer to a malicious query $q$. The goal of an inferential adversary \textsc{AdvLLM} with prior $p_{adv}(a|q)$ is to select $k$ inputs $\{q_i\}_{i=1}^{k}$ to maximize the scoring function
\begin{align*}
    s(q,h^k) =  \sum_{a \in A_q} p_{adv}(a|h^k,q)\log \frac{p_{adv}(a|h^k,q)}{p_{adv}(a|q)}.
\end{align*} 
where $h^k = \{(q_i,a_i)\}_{i=1}^{k}$ is the  knowledge pile of $k$ interactions with \textsc{VicLLM}.
\end{definition}

\section{Information Censorship}\label{sec:Censorship}
\paragraph{Overview} To mitigate risks incurred by inferential adversaries, the victim model provider seeks to minimize the expected impermissible information leakage over collections of interactions. In this section we define information censorship, a bound on expected impermissible information leakage (Exp-\textsc{IIL}) which defense mechanisms must ensure to guarantee safety against inferential adversaries. We propose a randomised response defense mechanism for ensuring this bound. 
 
\subsection{Safety Guarantee}

\begin{definition}[Expected Impermissible Information Leakage (Exp-\textsc{IIL})]
Let $M$ be a censorship mechanism such that for an input $q_i$ to the \textsc{VicLLM}, the answer $a_i$ returned to \textsc{AdvLLM} is sampled from $M(q_s,\textsc{VicLLM}(q_s))$. Assuming the \textsc{AdvLLM} seeks to answer $q \in Q$ and submits a set of $k$ queries $\{q_i\}_{i=1}^{k}$ to the \textsc{VicLLM}, define the distribution of knowledge piles $h^k = \{(q_i,a_i)\}_{i=1}^{k}$ to be $H_{q_i}^k := \{q_i,M(q_i,\textsc{VicLLM}(q_i))\}_{i=1}^{k}$. Then, the expected impermissible information leakage is given by:
\begin{align*}
    I_{A_q}(p_{adv}(\cdot|q);H_{q_i}^k) 
    = \sum_{\{a_1,\dots,a_k\} \in \mathcal{Y}^k}p_M(H_{q_i}^k = h^k)\sum_{a \in A_q}p_{adv}(a|h^k,q)
    \log \frac{p_{adv}(a|h^k,q)}{p_{adv}(a|q)}.
\end{align*}
\end{definition}

\myparagraph{Distinction from MI} Our definition of Exp-\textsc{IIL} differs from the mutual information definition
\begin{align*}
    I(p_{adv}(\cdot|q);H_{q_i}^k) 
    = \sum_{\{a_1,\dots,a_k\} \in \mathcal{Y}^k}p_M(H_{q_i}^{k}=h^k)  \sum_{a \in \mathcal{Y}}p_{adv}(a|h^k,q)
    \log \frac{p_{adv}(a|h^k,q)}{p_{adv}(a|q)}.
\end{align*}
This distinction arises as our intent is to capture the asymmetry regarding which answers $a$ an adversary becomes more confident in. From a safety perspective, the only concern is whether or not the adversary becomes more confident in \textit{impermissible} conclusions $a \in A_q$. Specifically, for the LLM provider, scenarios in which the adversary's posterior $p_{adv}(a|h^k,q) = 1$ for some $a \not \in A_q$ is perfectly acceptable as it implies the adversary is confident in a ``permissible'' answer to $q$, whereas a bound on mutual information would deem this a defense failure assuming the entropy of the adversary prior distribution $p_{adv}(\cdot|q)$ was high.

In order to defend against an inferential adversary, \textsc{AdvLLM}, from inferring the answer to a harmful query $q \in Q$ over $k$ interactions, a censorship mechanism $M$ must bound the worst case Exp-\textsc{IIL}.

\begin{definition}[$(k,\epsilon)$-Information Censorship Mechanism (ICM)]
For a collection of adversary priors $\Phi_{adv}$, a malicious query $q \in Q$, a leakage bound $\epsilon > 0$, and $k$ possible interactions between \textsc{AdvLLM} and \textsc{VicLLM}, a $(k,\epsilon)$-ICM $M$ ensures the worst-case Exp-\textsc{IIL} is bounded by $\epsilon$ for knowledge pile distribution $H_{q_i}^k := \{q_i,M(q_i,\textsc{VicLLM}(q_i))\}_{i=1}^{k}$:
\begin{align*}
\sup_{\substack{p_{adv} \in \Phi; \\ \{q_i\}_{i=1}^{k} \in \mathcal{X}^k}} I_{A_q}(p_{adv}(\cdot|q);H_{q_i}^k) \leq \epsilon.
\end{align*}

\end{definition} 

\myparagraph{Compositional Bounds} Finding and bounding the supremum of the Exp-\textsc{IIL} necessary for a $(k,\epsilon)$-ICM over all possible sets of $k$ interactions is increasingly challenging due to the combinatorial complexity of checking all combinations. However, the $(1,\epsilon)$-ICM (henceforth referred to as an $\epsilon$-ICM) can also provide bounds on Exp-\textsc{IIL} for $k$ interactions. 
Assuming the \textsc{AdvLLM} can interact with \textsc{VicLLM} across independent context windows, the $\epsilon$-ICM cannot depend on knowledge of an existing interaction history $h^i$---it must be non-adaptive. We provide a non-adaptive composition bound of an $\epsilon$-ICM inspired by results in \cite{nuradha2023pufferfish}. 

\begin{theorem}[Non-Adaptive Composability of $\epsilon$-ICM]\label{thm:compositional_bound}
For a collection of adversary priors $\Phi$, malicious query $q \in Q$, leakage bound $\epsilon > 0$, $k$ possible interactions between \textsc{AdvLLM} and \textsc{VicLLM}, and an $\epsilon$-ICM $M$, 
\begin{align*}
    \sup_{\substack{p_{adv} \in \Phi; \\ \{q_i\}_{i=1}^{k} \in \mathcal{X}^k}} I_{A_q}(p_{adv}(\cdot|q);H_{q_i}^k) \leq k\epsilon +  \sum_{j=2}^{k} I_{A_q}((q_j,a_j); H_{q_i}^{j-1}|p_{adv}(\cdot|q)).
\end{align*}
\end{theorem}

\myparagraph{Interpretation} In other words, the joint leakage can be bounded by the sum of $k$ individual $\epsilon$ per-interaction leakages and a term capturing the dependencies between interactions when conditioned on $p_{adv}(\cdot|q)$. If the model outputs $\textsc{VicLLM}(q_i)$ are deterministic or independent when conditioned on $p_{adv}(\cdot|q)$, then, the sum becomes $0$ because the noise mechanism for an $\epsilon$-ICM is independent of the response. Such assumptions could hold when there is a ``single true value'' of $p_{adv}(\cdot|q)$ known by \textsc{VicLLM}, and all model outputs are related to this value by a deterministic function. 

\subsection{Randomised Response $\epsilon$-ICM}
\myparagraph{Proposed Defense} To provide concrete bounds on the information leakage to a $k$-inferential adversary and demonstrate what an $\epsilon$-ICM could look like, we construct an $\epsilon$-ICM. Inspired by a differentially private mechanism proposed by \citet{mangat1994improved} to protect privacy of individuals during surveys, we propose a randomized response information censorship mechanism.

\begin{definition}[Randomised Response Mechanism]
Let $q \in [0,1]$ and $S \subset \mathcal{Y}$ be a nonempty set of safe strings such as the empty string. The randomised response mechanism $M_q: \mathcal{X}\times P(\mathcal{Y}) \to P(\mathcal{Y})$ is defined as:
    \begin{align*}
    p_M((q,a)) = \begin{cases}
        tp(\textsc{VicLLM}(q)=a) & \text{if } a \in \mathcal{Y}\setminus S
        \\ (1-t)\frac{1}{|S|} & \text{if } a \in S
    \end{cases}
\end{align*}
where $p_{\textsc{VicLLM}}(y)$ is the probability distribution of the victim model's output.
\end{definition}

\begin{theorem}[Randomised Response $\epsilon$-ICM]\label{thm:random_response_icm}
Let $\Phi$ be a collection of adversary priors. We assume there exists a nonempty safety set $S \subset \mathcal{Y}$ such that for any $q \in Q$, $p_{adv} \in \Phi$, and $s \in S$, $I_{A_q}(p_{adv}(\cdot|q); (q_1,s)) = 0$. If we let
\begin{align*}
    t_{\epsilon} = \min\left(\frac{\epsilon}{\sup_{\substack{p_{adv} \in \Phi; \\ q_1 \in \mathcal{X}}} I_{A_q}(p_{adv}(\cdot|q);H_{q_i}^1)}, 1\right)
\end{align*}
then the randomised response mechanism $M_{q_{\epsilon}}$ is an $\epsilon$-ICM.
\end{theorem}

In order to satisfy the non-adaptive compositional bounds, it becomes evident that the probability of returning the model response can become quite low, thereby affecting utility for benign users. We now turn to establishing these tradeoff results.

\subsection{Safety-Utility Trade-offs}\label{sec:safety-utility} 
\myparagraph{Defnining Utility} While the primary concern of censorship is to ensure safety by mitigating Exp-\textsc{IIL}, model providers also care about the utility of their model for benign users. Given a distribution $P_{\mathcal{X}}$ for the probability that a benign user prompts the \textsc{VicLLM} with $x$ we can define the utility of an interaction between the \textsc{VicLLM} and the benign user as: 
\begin{align*}
    \mathbb{E}_{y \sim \textsc{VicLLM}(x)}\left[ u(x,y) \right].
\end{align*}

Assuming that the safe responses $y \in S$ provide no utility for benign users, we easily find the utility cost of the Randomised Response $\epsilon-ICM$
\begin{theorem}[Utility Bound for Randomised Response $\epsilon$-ICM]\label{thm:rand_resp_utility_loss}
For a given input $x$ and utility function $u(x,y)$, the expected utility of an interaction where the outputs are given by the randomized response $\epsilon$-ICM can be bounded as follows:
\begin{equation}
\frac{\mathbb{E}_{y \sim M(x,\textsc{VicLLM}(x))}[u(x,y)]}{\mathbb{E}_{y \sim \textsc{VicLLM}(x)}[u(x,y)]} = q_{\epsilon},
\end{equation}
$t_{\epsilon}$ is the probability of the mechanism returning a response from the Victim model.
\end{theorem}
Thus, introducing the randomised response mechanism yields a model with utility $q_{\epsilon}$ that of the uncensored model for any inputs for which the mechanism is employed. 

\myparagraph{Inferential Users} If we assume benign users are also inferential, i.e. they seek to learn information from model outputs about some task of quesiton $x^*$, then, for prior $p_{usr}(\cdot|x^*)$, we can define their expected utility as:
\begin{align*}
    I(p_{usr}(\cdot|x^*);(x,Y)) = \sum_{y \in Y}p(x,y) \sum_{a \in \mathcal{Y}} p_{usr}(a|(x,y),x^*)\log \frac{p_{usr}(a|(x,y),x^*)}{p_{usr}(a|x^*)}.
\end{align*}

For such inferential users, an upper bound on the possible utility for any input $x$ can be derived for all possible $\epsilon$-ICMs. We adapt results from \cite{zamani2023privacy} to upper bound the mutual information.

\begin{theorem}[Utility Bound for Inferential Users]\label{thm:general_utility_loss} %
For a benign user interested in query $x^*$ with prior $p_{usr}(\cdot|x^*)$, a collection of adversary priors $\Phi$, set of malicious queries $Q \subset \mathcal{X}$,  with the \textsc{VicLLM} employing an $\epsilon$-ICM $M$, the utility for the benign inferential user interacting with the \textsc{VicLLM} employing and $\epsilon$-ICM, is bounded by:
\begin{align*}
I(p_{usr}(\cdot|x^*);(x,Y) \leq  \epsilon + \inf_{\substack{q \in Q \\ p_{adv} \in \Phi}}\left( \mathcal{H}(p_{usr}(\cdot|x^*)|p_{adv}(\cdot|q)) + I_{A_q^c}(p_{adv}(\cdot|q);(x,Y))\right) 
\end{align*}
where $A_q^c$ the complement of $A_q$ and $Y = M(x,\textsc{VicLLM}(x))$.
\end{theorem}

\myparagraph{Interpretation} Thus, the utility can only be high if the useful knowledge $p_{usr}(\cdot|x^*)$ is not specified by $p_{adv}(\cdot|q)$, or, the interaction is such that the mechanism $M$ makes $I_{A_q^c}(p_{adv}(\cdot|q);(x,Y))$ high, in other words making the adversary more confident in a permissible belief. While $\mathcal{H}(p_{usr}(\cdot|x^*)|p_{adv}(\cdot|q))$ does not depend on the mechanism employed, an $\epsilon$-ICM may be able to provide more utility if the responses returned make adversaries more confident in permissible conclusions.

\vspace{1em} %
\section{Related work and Discussion}
\subsection{Comparison to Jailbreaks}
The fundamental distinction between decomposition attacks and jailbreaks is the adversary's objective. While DAs share resemblance to methods such as payload splitting \citep{kang2023exploiting,li2024drattack} or multi-turn attacks \citep{russinovich2024great, li2024llmdefensesrobustmultiturn}, these methods still rely on a victim model to produce problematic outputs; output censorship methods could greatly limit their effectiveness. 

For example, payload splitting is an input obfuscation method which splits up harmful words or strings into substrings which are then assigned to symbolic variables and an equation containing these variables. The model is then asked the prompt, with
equations substituted for the sensitive words. Thus, payload splitting utilizes the programmatic capabilities of LLMs in order to mitigate the ability to detect harmfulness or malicious intent in the input, however, the outputs provided remain impermissible and open to output censorship mechanisms. 

Multi-turn jailbreaks \citep{russinovich2024great,li2024llmdefensesrobustmultiturn} bypass defenses by gradual escalation from benign prompt and response interactions to problematic ones. These attacks are related to many-shot jailbreaks \citep{anil2024many}, leveraging an extended context window to alter model behavior. Nevertheless, once again, the method aims to force the victim model to produce a strictly impermissible output, which could more easily be blocked by output filtering mechanisms and may not even assist the adversary in fulfilling their malicious goal.

\subsection{Relating inferential adversaries to privacy literature}
Inferential adversaries have been studied before in the context of privacy, such as model stealing adversaries \citep{tramer2016stealing} and membership inference adversaries \citep{shokri2017membership}. Model stealing adversaries utilize carefully crafted queries and certain compositional properties of these queries and their outputs to infer specific parameters \citep{shamir2023polynomial}. The black-box \textit{search} for adversarial examples can also be viewed as being performed by an inferential adversary, as the input queries are not directly causing a model failure but leak information about gradients \citep{ilyas2018black}. %

By presenting inferential adversaries as a compelling threat model in the context of AI safety, we establish parallels between the problems of AI safety and privacy. Initial attempts to address concerns of privacy in data often focused on direct anonymization, such as removing personally identifiable information (PII) from datasets \citep{10.1142/S0218488502001648}. These approaches operated under the assumption that by simply removing identifiers and forcing indistinguishability within a dataset would be sufficient for making individuals in the dataset unidentifiable \citep{sweeney2002k}. However, such approaches were insufficient for actually ensuring privacy because they did not account for the compositional nature of data. It was found that individuals could still be identified when certain combinations of attributes were unique especially given background information provided by auxiliary sources \citep{10.1145/1401890.1401926, 4531148}. Similarly, in the context of LLM safety, we argue that simply censoring or filtering direct responses does not address the underlying issue that an adversary can piece together sensitive information as part of a compositional attack. 

To address compositional attacks, Differential Privacy was proposed as a mathematical framework which defined and sought to bound the marginal risk to an individual's privacy when releasing data \citep{10.1007/11787006_1}. These guarantees are achieved by adding a controlled amount of noise to the data or to responses made to queries so as to ensure that any single individual's data does not significantly influence the outcome. Further definitions sought to generalize and extend privacy definitions in terms of information leakage \citep{nuradha2023pufferfish,grosse2024quantifying,bloch2021overview}. 

We hope that through introducing the inferential adversary threat model in the context of safety, we can inspire a paradigm shift similar to that which occurred for privacy for the field of AI safety. By showing that safety concerns can be similarly framed and that existing defense approaches are vulnerable to composition attacks leveraging multiple interactions and background knowledge, our work helps introduce a theoretically rigorous way for further assessing and establishing safety. %

\section{Conclusion}
We identify a key safety vulnerability of deployed LLMs in terms of impermissible information leakage. We contrast with existing work which treats the ability of LLMs to provide harmful information to adversaries as a problem of model robustness, highlighting that this perspective is too narrow. Although current robustness issues make it easy for security adversaries to get a desired result, as defenses improve such direct attacks may prove more difficult. However, \textit{robustness provides a false sense of safety}. The fundamental issue is that knowledge is compositional, interdependent, and dual-use. Directly censoring certain responses to questions does not guarantee that an adversary cannot reconstruct impermissible knowledge. Consequently, we conclude that evaluation methods and defenses must consider the potential for model responses to help adversaries infer dangerous information rather than determine whether or not model outputs themselves are dangerous. Nevertheless, we highlight that proper defenses will inevitably sacrifice utility for benign users.

\section{Acknowledgements}
We want to thank Cleverhans lab members Sierra Wyllie, Anvith Thudi, Mohammad Yaghini, and Stephan Rabanser, as well as Claas Voelcker, Roger Grosse, Erik Jones, Lev McKinney, Xander Davies, and Darija Barak for their helpful feedback. 

We would like to acknowledge our sponsors, who support our research with financial and in-kind contributions:
Amazon, Apple, CIFAR through the Canada CIFAR AI Chair, Meta, NSERC through the Discovery Grant and an Alliance Grant with ServiceNow and DRDC,
the Ontario Early Researcher Award, the Schmidt Sciences foundation through the AI2050 Early Career Fellow program, and the Sloan
Foundation. Resources used
in preparing this research were provided, in part, by the Province of Ontario, the Government of Canada
through CIFAR, and companies sponsoring the Vector Institute.

\newpage

\bibliography{example_paper}
\bibliographystyle{abbrvnat}  

\newpage
\appendix
\onecolumn
\section{Proofs}
As a few preliminaries for proving our results, we re-derive properties of expected impermissible information gain that match those of Mutual Information. For an $q \in Q$, we define a variant of Entropy
\begin{align*}
    \mathcal{H}_{A_q}(p_{adv}(\cdot|q)) = -\sum_{a \in A_q}p_{adv}(a|q)\log p_{adv}(a|q)
\end{align*}
and Conditional Entropy, conditioned on a knowledge pile distribution $H^k_{q_i}$ provided subquestions $\{q_i\}_{i=1}^k$
\begin{align*}
     \mathcal{H}_{A_q}(p_{adv}(\cdot|q) | H^k_{q_i}) = -\sum_{\{a_1,\dots,a_k\} \in \mathcal{Y}^k}p_M(H^k_{q_i}=h^k)\sum_{a \in A_q}p_{adv}(a|q,h^k)\log p_{adv}(a|q,h^k)
\end{align*}
resulting in the standard relationship $I_{A_q}(p_{adv}(\cdot|q); H^k_{q_i}) = \mathcal{H}_{A_q}(p_{adv}(\cdot|q) - \mathcal{H}_{A_q}(p_{adv}(\cdot|q)| H^k_{q_i})$.

\textbf{Non-negativity of $I_{A_q}(p_{adv}(\cdot|q); H^k_{q_i})$:} Denoting $p(A_q) = \sum_{a \in A_q} p_{adv}(a|q)$, Jensen's inequality gives us that
\begin{align*}
    I_{A_q}(p_{adv}(\cdot|q); H^k_{q_i}) = -p(A_q)(\sum_{a \in A_q}\sum_{h^k \in \mathcal{Y}^k}\frac{p_{joint}(a,h^k|q)}{p(A_q)}\log\frac{p_{adv}(a|q)p_M(H^k_{q_i}=h^k)}{p_{joint}(a,h^k|q)})
    \\ \geq -p(A_q)(\log(\sum_{a \in A_q}\sum_{h^k \in \mathcal{Y}^k}\frac{p_{joint}(a,h^k|q)}{p(A_q)}\frac{p_{adv}(a|q)p_M(H^k_{q_i}=h^k)}{p_{joint}(a,h^k|q)}))
    \\ = -p(A_q)\log(\sum_{a \in A_q}\sum_{h^k \in \mathcal{Y}^k}\frac{p_{adv}(a|q)p_M(H^k_{q_i}=h^k)}{p(A_q)})
    \\ = -p(A_q)\log(\sum_{a \in A_q}\frac{p_{adv}(a|q)}{p(A_q)})
    \\ = -p(A_q)\log(1) = 0
\end{align*}
where $p_{joint}$ is the joint distribution over adversary beliefs and knowledge piles.

\textbf{Chain rule of Impermissible Information:} We abstract a bit away from the specific distributions examined to define a chain rule for impermissible information. Specifically, given $q \in Q$, and $A_q \in \mathcal{Y}$, let $A$, $B$, and $C$ be random variables. Then, $I_{A_q}(C; A,B) = I_{A_q}(C;A) + I_{A_q}(C;B|A)$

The LHS can be expressed as 
\begin{align*}
    I_{A_q}(C; A,B) = \sum_{c\in A_q}\sum_{a,b} p(a,b,c) \log \frac{p(a,b,c)}{p(c)p(a,b)}
\end{align*}
whereas on the RHS
\begin{align*}
    I_{A_q}(C;A) = \sum_{c \in A_q} \sum_{a} p(a,c)\log \frac{p(a,c)}{p(a)p(c)} 
\end{align*}
and
\begin{align*}
    I_{A_q}(C|A;B)
    \\= \sum_{c \in A_q} \sum_{a,b} p(c, a, b) \log \frac{p(c,a,b)p(a)}{p(c,a)p(b,a)}
    \\ =  \sum_{c \in A_q} \sum_{a,b} p(c, a, b) \log \frac{p(c,a,b)}{p(c,a)p(b|a)}
\end{align*}
Thus, 
\begin{align*}
     I_{A_q}(C;A) + I_{A_q}(C;B|A) 
     \\ = \sum_{c \in A_q}\sum_{a, b} p(c, a, b) \left(\log \frac{p(c,a)}{p(c)p(a)} +  \log \frac{p(c, a, b)}{p(c, a)p(b | a)}\right)
     \\ =  \sum_{c\in A_q}\sum_{a, b} p(c, a, b)\log \frac{p(c,a,b)}{p(b|a)p(c)p(a)}
     \\ = \sum_{c\in A_q}\sum_{a, b} p(c,a,b)\log \frac{p(c,a,b)}{p(c)p(a,b)} 
     \\ = I_{A_q}(C; A,B)
\end{align*}
establishing the chain rule for our variant of mutual information. This argument can be further generalized by induction to provide 
\begin{align*}
    I_{A_q}(C; X_1,\dots,X_n) = \sum_{i=1}^{n}I_{A_q}(C|X_{1},\dots,X_{i-1}; X_i)
\end{align*}

\textbf{Data Processing Inequality:}
If an adversary applies some post processing to the knowledge pile signal $H^k_{q_i}$ to get $H^{k'}_{q_i}$, then $H^{k'}_{q_i}$ is conditionally independent of $p_{adv}(\cdot|q)$ given $H^k_{q_i}$. From the chain rule, we have that 
\begin{align*}
    I_{A_q}(p_{adv}(\cdot|q);H^k_{q_i},H^{k'}_{q_i}) = I_{A_q}(p_{adv}(\cdot|q);H^k_{q_i}) + I_{A_q}(p_{adv}(\cdot|q);H^{k'}_{q_i}|H^k_{q_i})
\end{align*}
and 
\begin{align*}
    I_{A_q}(p_{adv}(\cdot|q);H^k_{q_i},H^{k'}_{q_i}) = I_{A_q}(p_{adv}(\cdot|q);H^{k'}_{q_i}) + I_{A_q}(p_{adv}(\cdot|q);H^k_{q_i}|H^{k'}_{q_i})
\end{align*}
Conditional independence implies that $I_{A_q}(p_{adv}(\cdot|q);H^{k'}_{q_i}|H^k_{q_i}) = 0$, and, non-negativity implies that $I_{A_q}(p_{adv}(\cdot|q);H^{k'}_{q_i}|H^{k'}_{q_i})$, thus, $I_{A_q}(p_{adv}(\cdot|q); H^k_{q_i}) \geq I_{A_q}(p_{adv}(\cdot|q); H^{k'}_{q_i})$

\subsection{Proof of~\autoref{thm:compositional_bound}}
\begin{proof}
Let $q$ be a malicious query, $\epsilon > 0$ a leakage bound, and $k$ the number of possible interactions between \textsc{AdvLLM} and \textsc{VicLLM} mediated by an $\epsilon$-ICM $M$. For ease of notation, let $y_i = (q_i, M(q_i,\textsc{VicLLM}(q_i))$. For any set of $k$ questions $\{q_i\}_{i=1}^{k}$, and history distribution $H_{q_i}^{k} := \{q_i,a_i)\}_{i=1}^{k}$, we prove by induction that 
\begin{align*}
    I_{A_q}(p_{adv}(\cdot|q); H_{q_i}^{k}) \leq \sum_{i=1}^{k}\epsilon_i + \sum_{j=2}^{k} I_{A_q}(y_j;H_{q_i}^{j-1}|p_{adv}(\cdot|q))
\end{align*}
when $I_{A_q}(p_{adv}(\cdot|q); (q_j,a_j)) \leq \epsilon$.
\begin{align*}
    I_{A_q}((q_j,a_j);H_{q_i}^{j-1}|p_{adv}(\cdot|q))  = \\ \sum_{a \in A_q} p_{adv}(a|q)
    \sum_{y_1,\dots,y_j \in \mathcal{Y}}p_{M}(y_1,\dots,y_j|(q,a))\log \frac{p_{M}(y_1,\dots,y_j|(q,a))}{p_{M}(y_1,\dots,y_{j-1}|(q,a))p_{M}(y_j|(q,a))}
\end{align*}
For $k=1$, we trivially have that 
$I_{A_q}(p_{adv}(\cdot|q);H_{q_1}^{1}) \leq \epsilon \leq \epsilon + \eta$ as $\eta$ is non-negative.

Assume then that
\begin{align*}
    I_{A_q}(p_{adv}(\cdot|q); H_{q_i}^m) \leq m\epsilon + \eta_m
\end{align*}
for $\eta_m = \sum_{j=2}^{m}I_{A_q}(y_j;H_{q_i}^{j-1}|p_{adv}(\cdot|q))$. Then,
\begin{align*}
    I_{A_q}(p_{adv}(\cdot|q); H_{q_{i}}^{m+1}) 
    = I_{A_q}(p_{adv}(\cdot|q); H_{q_i}^m) + I_{A_q}(p_{adv}(\cdot|q);(q_{m+1},a_{m+1}|H_{q_i}^m)
    \\ \leq \sum_{i=1}^{m}\epsilon_i + \eta_m + I_{A_q}(p_{adv}(\cdot|q);(q_{m+1},a_{m+1}|H_{q_i}^m)
\end{align*}
To show our desired result, we need to show that 
\begin{align*}
    I_{A_q}(p_{adv}(\cdot|q);y_{m+1}|H_{q_i}^m) \leq I_{A_q}(p_{adv}(\cdot|q);y_{m+1}) + I_{A_q}(y_{m+1};H_{q_i}^{m}|p_{adv}(\cdot|q))
\end{align*}

For $I_{A_q}(p_{adv}(\cdot|q);y_{m+1}|H_{q_i}^m)$:
\begin{align*}
    I_{A_q}(p_{adv}(\cdot|q);y_{m+1}|H_{q_i}^m) =\\
    \sum_{h^m,a,y_{m+1}} p_{joint}(h^m,a,y_{m+1}|q) \log \frac{p_{joint}(h^m,a,y_{m+1}|q)p_{M}(h^m)}{p_{joint}(h^m,a|q)p_M(h^{m+1})}
\end{align*}

For $I_{A_q}(y_{m+1};H_{q_i}^m|p_{adv}(\cdot|q))$:
\begin{align*}
    I_{A_q}(y_{m+1};H_{q_i}^m|p_{adv}(\cdot|q)) =\\
    \sum_{h^m,a,y_{m+1}} p_{joint}(h^m,a,y_{m+1}|q) \log \frac{p_{joint}(h^m,a,y_{m+1}|q)p_{adv}(a|q)}{p_{joint}(h^m,a|q)p_{joint}(y_{m+1},a|q)}
\end{align*}

For $I_{A_q}(p_{adv}(\cdot|q);y_{m+1})$:
\begin{align*}
    I_{A_q}(p_{adv}(\cdot|q);y_{m+1}) = \\
    \sum_{h^m,a,y_{m+1}} p_{joint}(h^m,a,y_{m+1}|q) \log \frac{p_{joint}(a,y_{m+1}|q)}{p_{adv}(a|q)p_M(y_{m+1})}
\end{align*}

Thus, 
\begin{align*}
    I_{A_q}(p_{adv}(\cdot|q);y_{m+1}) + I_{A_q}(y_{m+1};h^{m}|p_{adv}(\cdot|q))
    \\= \sum_{h^m,a,y_{m+1}} p_{joint}(h^m,a,y_{m+1}|q) \log [\frac{p_{joint}(h^m,a,y_{m+1}|q)p_{adv}(a|q)}{p_{joint}(h^m,a|q)p_{joint}(y_{m+1},a|q)} \frac{p_{joint}(a,y_{m+1}|q)}{p_{adv}(a|q)p_M(y_{m+1})}]
    \\ = \sum_{h^m,a,y_{m+1}} p_{joint}(h^m,a,y_{m+1}|q) \log [\frac{p_{joint}(h^m,a,y_{m+1}|q)}{p_{joint}(h^m,a|q)p_{M}(y_{m+1})}]
\end{align*}
and 
\begin{align*}
    I_{A_q}(p_{adv}(\cdot|q);y_{m+1}) + I_{A_q}(y_{m+1};h^{m}|p_{adv}(\cdot|q)) - I_{A_q}(p_{adv}(\cdot|q);y_{m+1}|H_{q_i}^m) 
    \\ = \sum_{h^m,a,y_{m+1}} p_{joint}(h^m,a,y_{m+1}|q) \log [\frac{p_{joint}(h^m,a,y_{m+1}|q)}{p_{joint}(h^m,a|q)p_{M}(y_{m+1})}\frac{p_{joint}(h^m,a|q)p_M(h^{m+1})}{p_{joint}(h^m,a,y_{m+1}|q)p_{M}(h^m)}]
    \\ = \sum_{h^m,a,y_{m+1}} p_{joint}(h^m,a,y_{m+1}|q) \log [\frac{p_{M}(h^{m+1})}{p_{M}(y_{m+1})p_{M}(h^{m})}]
    \\ = \sum_{h^m} p_{M}(h^{m+1}) \log [\frac{p_{M}(h^{m+1})}{p_{M}(y_{m+1})p_{M}(h^{m})}]
    = I(y_{m+1};h^m) \geq 0
\end{align*}

Finally, we conclude that 
\begin{align*}
    I_{A_q}(p_{adv}(\cdot|q);y_{m+1}|H_{q_i}^m) \leq I_{A_q}(p_{adv}(\cdot|q);y_{m+1}) + I_{A_q}(y_{m+1};H_{q_i}^{m}|p_{adv}(\cdot|q)) \leq \epsilon + I_{A_q}(y_{m+1};H_{q_i}^{m}|p_{adv}(\cdot|q))
\end{align*} 
and $I_{A_q}(p_{adv}(\cdot|q); H_{q_i}^{m+1}) \leq (m+1)\epsilon + \eta_{m+1}$.

\end{proof}

\subsection{Proof of~\autoref{thm:random_response_icm}}
\begin{proof}
    Let $\Phi$ be a collection of adversary priors and $\epsilon > 0$. We assume there exists a nonempty safety set $S \subset \mathcal{Y}$ such that for any $q \in Q$, $p_{adv} \in \Phi$, and $s \in S$, $I_{A_q}(p_{adv}(\cdot|q); (q_1,s)) = 0$. Finally, let
\begin{align*}
    t_{\epsilon} = \min\left(\frac{\epsilon}{\sup_{\substack{p_{adv} \in \Phi; \\ q_1 \in \mathcal{X}}} I_{A_q}(p_{adv}(\cdot|q);H_{q_i}^1)}, 1\right)
\end{align*}
Then, for any input $q_1 \in \mathcal{X}$, we have 
\begin{align*}
    I_{A_q}(p_{adv}(\cdot|q);(q_1,H_{q_i}^1) = \sum_{\{a_1\} \in \mathcal{Y}}p_M((q_1,a_1)\sum_{a \in A_q}p_{adv}(a|a_1,q)
    \log \frac{p_{adv}(a|a_1,q)}{p_{adv}(a|q)} 
\end{align*}
If $a_1 \in S$, then by our assumption on $S$, $ \log \frac{p_{adv}(a|a_1,q)}{p_{adv}(a|q)} = 0$, thus,
\begin{align*}
    I_{A_q}(p_{adv}(\cdot|q);(q_1,H_{q_1}^1) = \sum_{\{a_1\} \in \mathcal{Y}\setminus S}p_M((q_1,a_1)\sum_{a \in A_q}p_{adv}(a|a_1,q)
    \log \frac{p_{adv}(a|a_1,q)}{p_{adv}(a|q)} 
    \\ = t_{\epsilon}\sum_{\{a_1\} \in \mathcal{Y}\setminus S}p(\textsc{VicLLM}(q_1)=a_1)\sum_{a \in A_q}p_{adv}(a|a_1,q)
    \log \frac{p_{adv}(a|a_1,q)}{p_{adv}(a|q)}
    \\ \leq t_{\epsilon} I_{A_q}(p_{adv}(\cdot|q);H_{q_i}^1)
    \\ \leq \epsilon
\end{align*}
\end{proof}
\subsection{Proof of~\autoref{thm:rand_resp_utility_loss}}

\begin{proof}
\begin{align*}
    &\mathbb{E}_{y \sim M(x,\textsc{VicLLM}(x))}[u(x,y)] = \sum_{y \in Y}p_{M}(y) u(x,y)
    \\& = t_{\epsilon}\sum_{y \in Y} p_M(y) u(x,y) + (1-t_{\epsilon})\mathbb{E}_{y \sim \text{Unif}(S)}[u(x,y)]
\end{align*}
Assuming $u(x,y) = 0$ for $y \in S$, we can express the utility ratio:
\begin{align*}
    \frac{\mathbb{E}_{y \sim M(x,\textsc{VicLLM}(x))}[u(x,y)]}{\mathbb{E}_{y \sim \textsc{VicLLM}(x)}[u(x,y)]} = \frac{t_{\epsilon} \sum_{y \in Y}p(y) u(x,y) + (1-t_{\epsilon})\cdot 0}{\sum_{y \in Y}p(y) u(x,y)} 
    = t_{\epsilon}
\end{align*}
\end{proof}

\subsection{Proof of \autoref{thm:general_utility_loss}}
\begin{proof}
By the chain rule for mutual information:
\begin{align*}
I(p_{usr}(\cdot|x^*);(x,Y)) &= I(p_{adv}(\cdot|q),p_{usr}(\cdot|x^*);(x,Y)) - I(p_{adv}(\cdot|q);(x,Y)|p_{usr}(\cdot|x^*)) \\
&= I(p_{adv}(\cdot|q);(x,Y)) + I(p_{usr}(\cdot|x^*);(x,Y)|p_{adv}(\cdot|q)) \\
&\quad - I(p_{adv}(\cdot|q);(x,Y)|p_{usr}(\cdot|x^*)) \\
&\leq I(p_{usr}(\cdot|x^*);(x,Y)|p_{adv}(\cdot|q)) + I(p_{adv}(\cdot|q);(x,Y))
\end{align*}

For any malicious query $q \in Q$ and adversary prior $p_{adv} \in \Phi$, we can decompose $I(p_{adv}(\cdot|q);(x,Y))$ into information gain about impermissible $A_q$ and permissible $A_q^c$ conclusions:

\begin{align*}
I(p_{adv}(\cdot|q);(x,Y)) = I_{A_q}(p_{adv}(\cdot|q);(x,Y)) + I_{A_q^c}(p_{adv}(\cdot|q);(x,Y)) \leq \epsilon + I_{A_q^c}(p_{adv}(\cdot|q);(x,Y))
\end{align*}

Using the fact that $I(p_{usr}(\cdot|x^*);(x,Y)|p_{adv}(\cdot|q)) \leq \mathcal{H}(p_{usr}(\cdot|x^*)|p_{adv}(\cdot|q))$, we get:

\begin{align*}
I(p_{usr}(\cdot|x^*);(x,Y)) &\leq \mathcal{H}(p_{usr}(\cdot|x^*)|p_{adv}(\cdot|q)) + I_{A_q^c}(p_{adv}(\cdot|q);(x,Y)) + \epsilon
\end{align*}

Since this holds for any $q \in Q$ and $p_{adv} \in \Phi$, we can take the infimum:

\begin{align*}
I(p_{usr}(\cdot|x^*);(x,Y)) \leq \epsilon + \inf_{\substack{q \in Q \\ p_{adv} \in \Phi}}\left(\mathcal{H}(p_{usr}(\cdot|x^*)|p_{adv}(\cdot|q)) + I_{A_q^c}(p_{adv}(\cdot|q);(x,Y))\right)
\end{align*}

\end{proof}

\section{Decomposition Attack Algorithm Implementation details}\label{sec:DA_deets}

\subsection{Experimental Details}
\myparagraph{Dataset Selection and Curation:} To evaluate DAs and compare them against jailbreaks under our proposed \textsc{IIL} metric, we chose the WMDP dataset \citep{li2024wmdp}, focusing on multiple-choice questions in biology and chemistry (WMDP-Bio and WMDP-Chem) as proxies for dangerous knowledge. To assess DAs' ability to extract impermissible information despite filtering mechanisms, we used Llama-Guard-3-8B \citep{dubey2024llama3herdmodels} to curate subsets of questions classified as unsafe, resulting in 35 and 64 questions for WMDP-Bio and WMDP-Chem, respectively. This curation ensures the all questions would trigger input filtering mechanisms if asked directly.

\begin{table}[h]
\centering
\begin{tabular}{ll}
\multicolumn{2}{c}{\textbf{Models Used}} \\[2ex]
\toprule
\textbf{Model Name} & \textbf{Purpose} \\
\midrule
Llama-3.1-70B-Instruct \citep{dubey2024llama3herdmodels} & Victim model ($\textsc{VicLLM}$) \\

Prompt-Guard-86M \citep{dubey2024llama3herdmodels} & Jailbreak detection for Input Filtering ($M_i$) \\

Llama-Guard-3-8B\citep{dubey2024llama3herdmodels} & Input and output filtering mechanism ($M_i$ and $M_o$) \\

Mistral-7B-Instruct-v0.3 \citep{jiang2023llm} & Adversary LLM (AdvLLM); not safety alignmed \\

Llama-3.1-8B-Instruct \citep{dubey2024llama3herdmodels} & Adversary LLM  (AdvLLM); safety alignmed \\
\bottomrule
\end{tabular}
\caption{Models and their purpose in the experiment. All sampling for generation is performed with temperature $t=0.8$ and nucleus sampling with $p=0.95$.}
\label{tab:models}
\end{table}

\myparagraph{Few-shot Examples:} Some few-shot examples for subquestion generation $E_g$ and answer aggregation $E_a$ were adapted from \citep{anthropicquestiondecomposition} which focused on problem decomposition and aggregation for problem solving. The remaining few shot examples were generated using Claude 3.5 Sonnet \citep{anthropicIntroducingClaude} to match the more technical multiple choice questions in the WMDP dataset. 

\myparagraph{Hyperparameters:} We run our proposed DA with $k=2$ rounds and $m=3$ subquestions per round. To reflect more realistic use cases and remove outliers, we only apply the attack to those questions for which the initial answer entropy is greater than $0.5$.\footnote{This implies that the adversary assigns less than about $.97$ probability to any one of the choices} For each question we run the DA $5$ times, returning the \textsc{IIL} measurements for each run as well as the average number of times inputs or outputs were flagged by the input or output filtering models per attack.

\appendix

\end{document}